# Hard-material Adhesion: Which Scales of Roughness Matter?

L. A. Thimons[1] · A. Gujrati[1] · A. Sanner[2,3] · L. Pastewka[2,3] · T. D. B. Jacobs[1]

**Abstract**

**Background** Surface topography strongly modifies adhesion of hard-material contacts, yet roughness of real surfaces typically exists over many length scales, and it is not clear which of these scales has the strongest effect. Objective: This investigation aims to determine which scales of topography have the strongest effect on macroscopic adhesion.

**Methods** Adhesion measurements were performed on technology-relevant diamond coatings of varying roughness using spherical ruby probes that are large enough (0.5-mm-diameter) to sample all length scales of topography. For each material, more than 2000 measurements of pull-off force were performed in order to investigate the magnitude and statistical distribution of adhesion. Using sphere-contact models, the roughness-dependent *effective* values of work of adhesion were measured, ranging from 0.08 to 7.15 mJ/m$^2$ across the four surfaces. The data was more accurately fit using numerical analysis, where an interaction potential was integrated over the AFM-measured topography of all contacting surfaces.

**Results** These calculations revealed that consideration of nanometer-scale plasticity in the materials was crucial for a good quantitative fit of the measurements, and the presence of such plasticity was confirmed with AFM measurements of the probe after testing. This analysis enabled the extraction of geometry-independent material parameters; the *intrinsic* work of adhesion between ruby and diamond was determined to be 46.3 mJ/m$^2$. The range of adhesion was 5.6 nm, which is longer than is typically assumed for atomic interactions, but is in agreement with other recent investigations. Finally, the numerical analysis was repeated for the same surfaces but this time with different length-scales of roughness included or filtered out.

**Conclusions** The results demonstrate a critical band of length-scales—between 43 nm and 1.8 µm in lateral size—that has the strongest effect on the total adhesive force for these hard, rough contacts.

## Introduction

All real surfaces exhibit roughness, which has profound effects on surface properties. This includes the mechanics of interfaces: adhesion [1, 2], contact stiffness [3–5], wetting [6], and friction [7]. Various analytical models have been developed to describe the dependence of functional properties on the geometry of the rough surface. The classic Greenwood and Williamson [8] multiasperity model for contact between rough surfaces was extended by Fuller and Tabor [1] and Maugis [9] to include adhesion. Further progress was made in connecting contact properties and roughness by Bush, Gibson, and Thomas (BGT) [10] and Rumpf [11, 12]. These models approximate real-world roughness using simpler mathematical functions and typically associate properties with a single geometric parameter, such as the root-mean square (RMS) height. More recent models have attempted to explicitly account for the hierarchical, multi-scale nature of roughness. Mandelbrot began characterizing surfaces as fractal-like and self-affine using spectral analysis [13]. Later, Persson [14] created a theory for rubber friction which draws quantitative connections between fractal roughness and contact properties, including adhesion [15]. These multi-scale models are expansions that start from the limit of conforming contacts. They balance the adhesive energy gained when making contact with the stored elastic deformation energy, and are expected to work well for soft, conformal contacts.

✉ T. D. B. Jacobs
tjacobs@pitt.edu

[1] Department of Mechanical Engineering and Materials Science, University of Pittsburgh, Pittsburgh, PA 15261, USA

[2] Department of Microsystems Engineering, University of Freiburg, Georges-Köhler-Allee 103, 79110 Freiburg, Germany

[3] Cluster of Excellence livMatS, Freiburg Center for Interactive Materials and Bioinspired Technologies, University of Freiburg, Georges-Köhler-Allee 105, 79110 Freiburg, Germany

A simpler approach is sufficient for hard contacts, in which the elastic energy required for creating a conforming contact is much larger than the interfacial adhesive energy gain: An interaction potential is integrated over the undeformed contacting geometry. This approach can be applied to simple analytical geometries, such as spheres [16, 17], and can also be applied to more complex geometries, including rough surfaces, sharp tips, etc. [18–20]. Pioneering work by Delrio et al. [18] showed that long-range Casimir forces contribute strongly to surface adhesion of ultra-flat (RMS height of 2–10 nm) micromachined surfaces at separation distances up to tens of nanometers.

Three critical questions remain for describing the roughness-dependent adhesion of surfaces with multi-scale topography. First, can the classic analytical models such as those proposed by Fuller and Tabor, Rumpf, and Maugis, be applied to describe their behavior? Second, is the importance of long-range adhesive interactions limited to ultra-flat surfaces (as found in [18]) or is it generalizable to real-world coatings that are rough over many length scales? Third, and most generally, which size-scales contribute most strongly to adhesion?

The purpose of the present work is to investigate these questions using adhesion tests of some of the hardest materials: a ruby sphere on polycrystalline diamond substrates. Diamond coatings are technologically relevant in a number of applications [21], including medical devices [22, 23], tool coating [24], face seals [25], and microelectromechanical systems (MEMS)[26]. The roughness of these films can be controlled by varying the growth condition or by polishing. Thus, we can test substrates with varying roughness but nominally identical surface chemistry. This allows us to isolate the effects of topography on adhesion. The surface topography [27] of these materials and their adhesion to soft PDMS [28] has been extensively characterized in prior publications. The present investigation examines their adhesion to a hard material: ruby.

Many adhesion studies have used atomic force microscopy (AFM) or colloidal AFM to characterize surface topography and then perform tip-based adhesion tests on the measured surface [29–32]. Such investigations provide valuable information on the atomic-scale parameters governing nanoscale adhesion. However, the small size of the contact limits their applicability in understanding the contribution of multi-scale roughness to macroscale adhesion. The present investigation overcomes this limitation by using AFM to characterize the topography, while using a large 0.5-mm-diameter sphere to measure adhesion.

## Methods

### Experimental Adhesion and Topography Measurements

Adhesion tests were carried out between ruby spheres and polycrystalline diamond coatings using a MEMS-based force sensing probe (FT-MA02, FemtoTools, Buchs, Switzerland). The 0.5-mm-diameter spheres (B0.50R, Swiss Jewel, Philadelphia, PA) were pre-polished to an ultra-smooth (RMS height < 1 nm) finish using a slurry of ruby particles (0.05 μm). The spheres were glued to the tips of the force-probes to create a sphere-on-flat geometry for the test. The substrates comprised four different polycrystalline diamond coatings, which were grown by hot-filament chemical vapor deposition (HF-CVD) and are boron-doped for electrical conductivity. The substrates have varying grain size, and are denoted microcrystalline diamond (MCD), nanocrystalline diamond (NCD), ultrananocrystalline diamond (UNCD), and a polished form of UNCD. The deposition and surface topography of the diamond coatings are characterized in Ref. [33].

For this work, the topography of the spheres and substrates were measured using atomic force microscopy (AFM) (Dimension V, Bruker, Billerica, MA). Measurements were made using diamond-like carbon (DLC)-coated probes (Tap DLC300, Mikromasch, Watsonville, CA) in tapping mode. Scans with lateral size of 2.5 μm (512×512 pixels) were performed on each of the four substrates and on the ruby spheres.

Representative images for the surface topography of the substrates, and of the polished spheres are shown in Fig. 1. Using the AFM measurements of 2.5-micron lateral size, the root-mean-square height, slope, and curvature of the surfaces are given in Table 1. The surface topography of the polycrystalline diamond films has been extensively measured in Ref. [27]. The numerical analysis included here, however, uses only the AFM-based topography measurements.

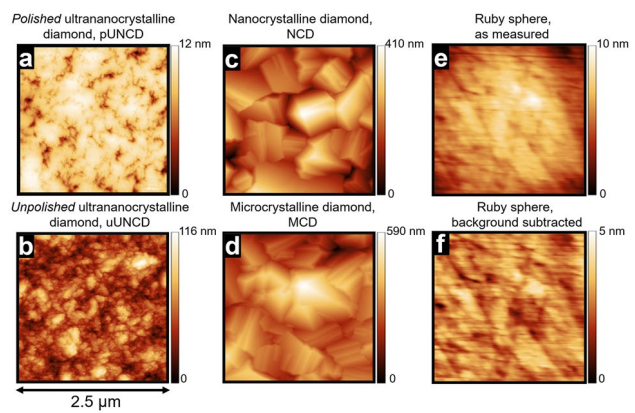

**Fig. 1** AFM measurements of the four polycrystalline diamond substrates (**a-d**) and one instance of a ruby sphere (**e**). The ruby sphere is also shown with its spherical geometry subtracted (**f**) to allow for direct comparison of roughness against the other substrates

The MCD and NCD have the largest roughness magnitude of the four substrates, as is shown by the vertical scale bar of the AFM images in Fig. 1. These two surfaces also show clear faceting due to the grain structure. The unpolished UNCD is significantly smoother than the MCD and NCD, and the faceting is not apparent at these size scales. The features that are visible have been shown to correspond to clusters of much smaller grains [27]. The polished UNCD is the smoothest of the four surfaces and shows no obvious grain structure. Similarly, the ruby tip images show a very smooth surface; while scratches are visible from the polishing process, the peak-to-valley roughness of this ruby sphere is smaller than all other surfaces. These AFM images, along with four more measurements from different sample areas of the various materials, form the basis of the numerical analysis that was performed.

The adhesion testing was performed using a custom micromechanical tester in a controlled-environment vacuum chamber on a vibration-isolation table. Dry air was flowed into the chamber prior to testing until the relative humidity was less than 1% (below the minimum reading of the humidity sensor). Dry air was flowed in for the duration of the test at low flow rates to ensure consistently low humidity levels.

A three-axis slip-stick piezoelectric stage provides closed-loop motion control and real-time x–y–z position data. For each individual adhesion measurement, the sphere was brought into contact with the substrate, loaded to a 10-μN preload (corresponding to a nominal Hertz stress of 135 MPa), and then withdrawn at a rate of 30 nm/s. The 10-μN preload occurs before the test and represents the minimum force required for the probe to find the point of contact. After finding contact the tip is lowered slowly onto the substrate up to a preload of 5μN.

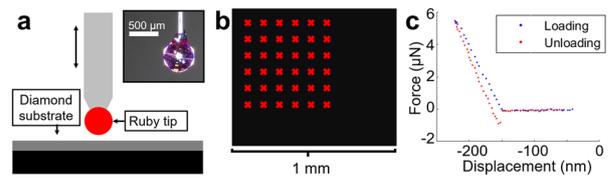

**Fig. 2** The tester consisted of a ruby sphere glued to a MEMS-based force probe, as shown schematically (**a**) and in an optical image (inset). Arrays of measurements (**b**) were performed on a single sample, with one measurement per location. A typical force–displacement curve is shown in (**c**), with the most-negative point recorded as the pull-off force

Then, the force required to pull the sphere off of the surface is recorded as $F_{pull-off}$, with a force resolution of 30 nN. A single "test" comprised an array of individual adhesion measurements (typically 1-by-1 mm), for a total of 400 individual measurements per test. Tests were performed in immediate sequence on all four samples in randomized order, without opening the chamber or modifying test conditions. Six such sequences were performed with different spheres and different substrate samples, to ensure repeatability and reproducibility of results. A schematic of a typical test setup is shown in Fig. 2.

## Numerical Analysis of Results

The experimental data was fitted using a cohesive zone model, using an exponential interaction potential with energy $U$ given by

$$U(r) = -W_{adh,int} \exp\left(\frac{-r}{\rho}\right) \tag{1}$$

with a hard-wall repulsion at $r = 0$. Here, $W_{adh,int}$ is the intrinsic work of adhesion, $r$ is the distance between interacting bodies, and $\rho$ is the "range of adhesion" [34], which describes the characteristic length scale of the adhesive interaction. Note that $U$ and $W$ are energies per unit surface area. Since we do not know the exact nature of the atomic-scale interaction between the two surfaces, this approach of using an empirical exponential interaction potential is a pragmatic approach to simplify the mathematical calculations. Similar results would be obtained using more complicated functional forms, e.g. based on instantaneous or retarded dispersion interactions, electrostatic interaction, or others. Distinguishing precisely between competing functional forms from our macroscopic experiments would be difficult. The functional form given in Eq. (1) allows separate fitting of the intrinsic work of adhesion ($U$ at distance $r = 0$) and the range of adhesion $\rho$, which yields the strength and length-scale of the interaction. This interaction potential has been widely used, including in the recent Contact Mechanics Challenge [34].

**Table 1** RMS roughness values calculated from AFM data only (top) and also from the full-spectrum roughness data, (bottom) which includes topography data from stylus, AFM, and TEM measurements of the same surfaces

| AFM Data Only | | Polished UNCD | Unpolished UNCD | NCD | MCD |
|---|---|---|---|---|---|
| | RMS height (nm) | $3.2 \pm 1.6$ | $17.4 \pm 2.1$ | $97.1 \pm 11.2$ | $107.1 \pm 12.0$ |
| | RMS slope | $0.04 \pm 0.01$ | $0.39 \pm 0.05$ | $0.51 \pm 0.07$ | $0.49 \pm 0.07$ |
| | RMS curvature (nm$^{-1}$) | $2.0 \times 10^{-3} \pm 1.3 \times 10^{-3}$ | $4.8 \times 10^{-2} \pm 3.8 \times 10^{-3}$ | $3.9 \times 10^{-2} \pm 1.1 \times 10^{-2}$ | $2.8 \times 10^{-2} \pm 6.6 \times 10^{-3}$ |
| Stylus, AFM & TEM Data [11] | | Polished UNCD | Unpolished UNCD | NCD | MCD |
| | RMS height (nm) | $4.2 \pm 0.8$ | $17.4 \pm 1.3$ | $97.2 \pm 11.7$ | $101.2 \pm 8.0$ |
| | RMS slope | $0.31 \pm 0.08$ | $1.17 \pm 0.28$ | $0.92 \pm 0.10$ | $0.85 \pm 0.10$ |
| | RMS curvature (nm$^{-1}$) | $1.99 \pm 0.35$ | $6.32 \pm 1.20$ | $5.91 \pm 1.83$ | $5.04 \pm 1.45$ |

This interaction potential can be converted into a cohesive law (stress-distance relationship) for two interacting bodies,

$$p(r) = -\frac{dU}{dr} = -\frac{W_{\text{adh,int}}}{\rho} \exp\left(\frac{-r}{\rho}\right) \qquad (2)$$

where $p$ is the (compressive) pressure acting between the surfaces. For the present analysis, this potential was applied to each pixel pairing between the substrate and the tip, resulting in the following calculated force $F_{\text{calc}}$ between the two contacting surfaces at separation $d$:

$$F_{\text{calc}}(d) = -\sum_{x,y} \frac{W_{\text{adh,int}}}{\rho} \exp\left(\frac{-(g_{x,y} + d)}{\rho}\right) A_{\text{pix}} \qquad (3)$$

Here, $A_{\text{pix}}$ is the area for a single surface pixel and the sum runs over all pixels in $x$ and $y$. Note that $g_{x,y}$ in Eq. (3) is the *difference* of the topography maps of the ruby sphere and the diamond coating, while $g_{x,y} + d$ is the gap between the two interacting surfaces. The calculation can only be carried out for distances $d$ where the gap $g_{x,y} + d$ is non-negative everywhere and the surfaces do not interpenetrate. The calculated adhesion values were thus found by summing the interaction potential pixel-by-pixel over every pixel pair of the two scans. The pull-off force is the minimum value of the force-separation curve $F_{\text{calc}}(d)$ that is found at the point of closest approach $d = $ -min $g_{x,y}$.

Due to random topography variation, there were sometimes significant contributions to adhesion from near the edges of the AFM scans. Therefore, for all substrate-tip combinations, the scans were stitched together so that there were no longer edge contributions to the adhesive interaction between the rough surface and a sphere of 0.5-mm diameter (<0.5% change from additional stitching). The stitching was done by mirroring the surface scans horizontally and vertically to ensure that all edges matched up. This was needed because real topography measurements are not periodic. Similar to the experiments, the tip was brought into contact with the substrate in many locations over a square array.

The above *rigid* analysis was supplemented by an elastic and a plastic analysis. For the elastic analysis, we computed surface deformation using a Fast-Fourier-transform-accelerated boundary element technique [2, 35]. The hard-wall constraint was realized using an L-BFGS-B optimizer [36]. Note that we do not report the results of the elastic analysis explicitly here, but it was carried out to rule out the influence of elastic surface deformation. From this elastic analysis, we generally found that the surface pressure was exceeding common hardness values in most of the contact area such that a purely plastic analysis is appropriate.

For the plastic analysis, we use a simple bearing-area approach. This assumes that the harder surface plastically deforms the softer surface on all points that penetrate, and that the pressure in the contact area is equal to the hardness $H$. The penetration of the tip is then such that the number of contacting (and hence plastically deformed) surface pixels is sufficient to support the preload, $N_{\text{contact}} = F/HA_{\text{pix}}$. A preload of 10 μN, chosen to match the experimental preload, was used to determine the amount of plastic deformation of the softer surface. No plastic deformation then occurs during pull-off; the pull-off force is simply a result of the plastically deformed geometry. Note that we did not employ a combination of elastic and plastic contact, but similar plasticity models were used in elasto-plastic contexts in Refs. [37–39]. For the plastic analysis, surfaces were brought into contact up to the specified preload, and the deformed surface at this preload was then used in a rigid pull-off calculation.

## Experimental Results

The topography can be used to compute roughness metrics such as root-mean-square (RMS) height, slope, and curvature, which are commonly used as inputs for rough-contact models. Table 1 shows the results of these calculations when performed only using the AFM measurements from this investigation (top) as compared to the same parameters that are computed when all of the many scales of roughness are included (bottom, using the full multi-scale spectral analysis from Ref. [27]). Table 1 shows the values that might serve as inputs to classical models, such as those of Maugis [9] or Rumpf [11]. Table 1 also serves to underscore just how widely varying these parameters can be when measured at different length scales. For example, the root-mean-square slope, a parameter that has been identified as important for multi-scale roughness models, varies for polished UNCD from 0.04 to 0.31 depending on how it is measured.

Figure 3 shows the distributions of values for adhesion force that have been measured on these four substrates. More than 2000 adhesion tests have been performed on every surface, with at least 6 different spheres. Each color on the histogram represents a new ruby sphere on a new sample of the diamond substrate. The mean adhesion of the polished UNCD was far higher than the unpolished version of the same material. Both UNCD surfaces showed higher adhesion than either the MCD or the NCD. The mean (median) pull-off forces of all four surfaces are 0.11 μN (0.06 μN), 0.16 μN (0.12 μN), 0.4 μN (0.31 μN), and 8.8 μN (1.75 μN), for MCD, NCD, unpolished UNCD, and polished UNCD respectively. Due to the large skew in the data, the mean value is shifted away from the peak of the distribution toward higher adhesion values.

All four of the distributions can be fit using a log-normal distribution. This is more difficult to see in the linear plots, given the skew of a large number of events with small pull-off force.

Plotted on a log–log scale (Fig. 3, bottom), the log-normal distribution is shown to accurately fit the data over at least two decades of adhesive force. In all cases, the low end of the distribution is cut off at 30 nN, as this represents the sensitivity of the force probe. Similar shapes for adhesion distributions have been reported previously for measurements in various contexts, including: centrifugal adhesion studies of particle adhesion in powders [40, 41]; biological samples and cell adhesion [42, 43] and many other studies using AFM adhesion measurements [44–46]. The origin of this distribution shape is not yet clear. While the fit is good for a log-normal distribution, there are other distributions—such as half of a gaussian distribution an inverse-gaussian distribution— that also give good qualitative fits. Further investigation is required to ascertain the origin of the shape of these distributions.

## Discussion

### Effective Work of Adhesion and the Application of Classical Rough-adhesion Models

A standard method of analyzing adhesion in rough spheres is to use classical sphere-contact models (such as JKR or DMT [47]) to extract an *effective* work of adhesion $W_{adh,eff}$, and then to use standard roughness models (such as those described in the first paragraph of the introduction) to relate $W_{adh,eff}$ to standard roughness parameters. Following the procedure of Grierson et al. [48], with material parameters of ruby (elastic modulus E = 365 GPa and Poisson ratio ν = 0.29) and diamond (E = 1010 GPa, ν = 0.22), a nominal sphere diameter of 0.5 mm, the Tabor parameter is determined to be 0.81. This falls in the transition region between the DMT and JKR models. Using Maugis' analysis for the transition region between JKR and DMT, and an approximate equilibrium spacing of 0.3 nm, the analysis yields values of $W_{adh,eff}$ = 0.08, 0.13, 0.32, and 7.15 mJ/m² for MCD, NCD, unpolished UNCD, and polished UNCD respectively. The surface chemistry is assumed to be similar for all of these HF-CVD diamond coatings, and therefore this difference is attributed primarily to surface topography.

It is clear from these measured values of effective work of adhesion, along with the values of RMS parameters shown in Table 1, that there are no simple relationships between RMS parameters and effective work of adhesion. Attempts to fit this data using simple analytical models [1, 9, 11] were unsuccessful, regardless of which roughness parameters were used (AFM-based or multi-scale). One potential explanation for why these models fail here is that the pull-off force for these hard materials is most dependent on the behavior of the uppermost contact points. These contacts represent the extreme-value statistics of the distribution of surface heights. They do not follow the central limit theorem

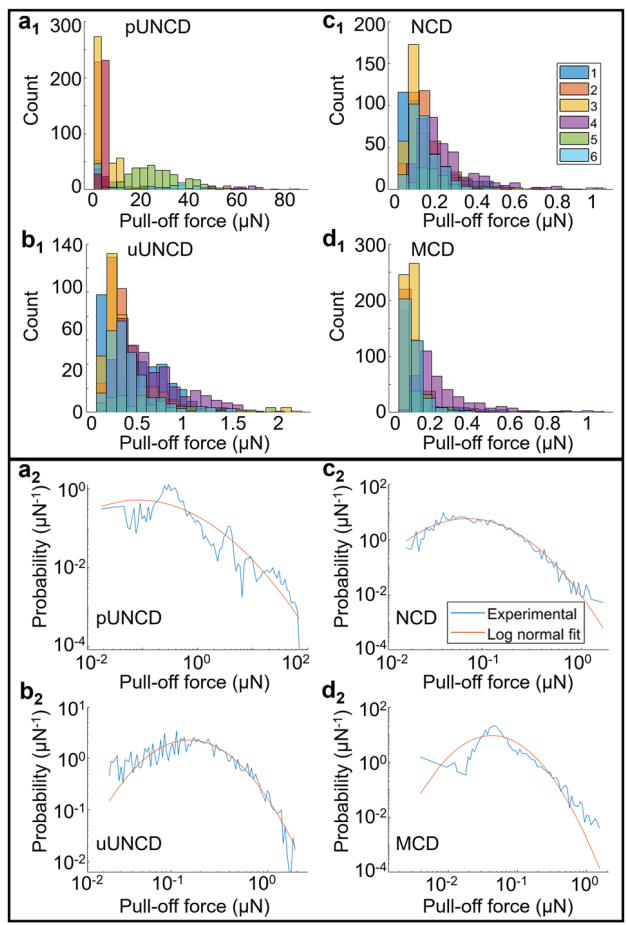

**Fig. 3** Experimental pull-off data is shown for the four substrates across different testing sessions (different colors) on a linear scale ($a_1$-$d_1$). The same data from the top panel is combined into single datasets and shown on log–log plots ($a_2$-$d_2$), with log-normal fits shown (red lines)

and are likely not described by the many models based on Gaussian statistics.

## Intrinsic Work of Adhesion and Range of Adhesion

Instead of a single-asperity model extracting the effective work of adhesion for each substrate, a numerical analysis can be performed using the combined roughness of the sphere and substrate (see Methods). Like the experiments, the calculations were repeated for an array of $20 \times 20$ contacts on each substrate. Each numerical calculation yields a computed pull-off force $F_{calc}$ for a specific choice of input values for $W_{adh,int}$ and $\rho$ and a specific contact location on the rough surface (Fig. 4a). Then a fitting routine can be applied to all data to extract the best-fit values of those material parameters.

For numerical tractability, the analysis was only performed over a square of size of $12.5\ \mu m$ rather than the 1-mm size scale of the experiments. Initially, the calculations were performed assuming rigid and/or elastic deformation only. Elastic calculations were virtually indistinguishable from the rigid calculations and we concluded that elasticity does not play a significant role in these contacts. Additionally, the micronewton-scale adhesive forces measured experimentally could not be explained with adhesion models based on rigid or elastically deforming surfaces, thus indicating that plastic deformation is likely occurring in these contacts.

We incorporated plastic deformation of the ruby tip with hardness of $H = 25$ GPa [49, 50] into our model using a penetration hardness model (see Methods). We first determine the plastic deformation in the softer sphere at a preload of $10\ \mu N$, identical to the average load used to find contact in the experiments. Pull-off calculations using the plastically deformed topography of the sphere were able to accurately reproduce the micronewton scale of pull-off forces from the experimental data. The possibility of plastic flow in similar hard materials has been reported in nanopillars [51] and nanoparticles [52]. A more in-depth analysis of the role of plasticity in these contacts is included in the following section.

The numerical analysis can be fit to the mean values of the experimental data from all four substrates in order to extract best-fit values for intrinsic work of adhesion and the range of adhesion. We note that while $W_{adh,int}$ simply rescales the computed pull-off force, the dependence of $F_{calc}$ on $\rho$ is nonlinear and depends on the specific topography. The range of adhesion $\rho$ is extracted by analyzing the *ratios* of pull-off forces between materials, since this cancels the (unknown) intrinsic work of adhesion $W_{adh,int}$ in our model equations. Figure 4b shows the ratio of the pull-off force of polished UNCD, NCD and MCD with respect to unpolished UNCD. The solid horizontal lines are the experimental results and the data points represent calculations carried out at various values of $\rho$ (x-axis). The error bars represent the variation over the contact points. Only for a range of adhesion of approximately $\rho = 5$ nm do all three lines cross the experimental results *simultaneously*. This means that while different values of $\rho$ (with modified values of $W_{adh,int}$) can describe individual experiments, a simultaneous fit yields a range of adhesion around 5 nm. Note that the increase in pull-off force for small values for range of adhesion is due to the finite pixel size. Once the range of adhesion was fit, the data was scaled by a factor $W_{adh,int}$ to match the magnitude of the experiments. A second relative error minimization was performed to find the best-fit value for $W_{adh,int}$ at the best-fit range of adhesion. Figure 4c shows the computed pull-off results calculated at various points on the rough topography as a function of range of adhesion $\rho$. The work of adhesion used in this plot is the value that yields the best possible final fit.

It is clear from Fig. 4 that the range of adhesion strongly affects the values of adhesion force. Rougher surfaces, like MCD and NCD, are less strongly affected and can be fit over a wider band of values for $\rho$. Smoother surfaces, such as the polished UNCD, are more influenced by changes in $\rho$ because the increasing range of adhesion enables more of the substrate to contribute to adhesion. This can be seen in Fig. 4c as a steeper slope for the smoothest polished UNCD surface and for the unpolished UNCD. The majority of the adhesion contribution to the rougher surfaces (MCD, NCD) comes from just one or two asperities, and therefore larger values for range of adhesion do not lead to such significant contributions to the area of interaction.

There is only one combination of parameters that enables the best fit for all samples. The fit was evaluated by computing and minimizing the mean relative error (MRE) between the fitted and measured adhesion. The uncertainty in the fit was computed for all values around the best-fit value with MRE < 0.1. This match between calculated and experimental data was used to extract values for work of adhesion of 46.3 ± 3.5 mJ/m² and range of adhesion of 5.6 ± 0.5 nm. Previous adhesion measurements on rough contacts between diamond and other hard materials report similar values for the intrinsic work of adhesion by accounting for surface roughness [20, 53, 54].

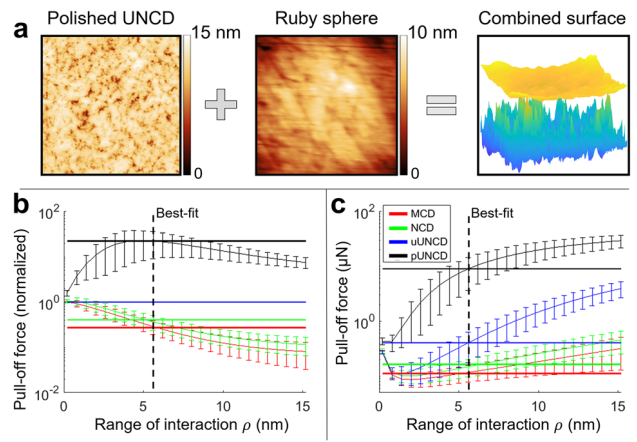

**Fig. 4** Computed pull-off forces were calculated by integrating an interaction potential over the combined roughness of the sphere and substrate (**a**). The best-fit value of range of adhesion was found by fitting to ratios of pull-off force (**b**), to eliminate the absolute value of work of adhesion. Finally, the absolute values of pull-off force were matched (**c**) by finding the best-fit value of intrinsic work of adhesion

The measured range of adhesion is much longer than is expected for typical atomic interactions such as covalent bonds or van der Waals forces, which are typically considered to have a range of adhesion around 0.3 to 0.6 nm [55]. However, there is prior nanoscale literature that supports a larger-than-expected value for range of adhesion. Using DLC-coated AFM tips, Grierson et al. have measured a range of adhesion between DLC and UNCD of 4–5 nm [56]. While for spherical (parabolic) tips the pull-off force does not depend on range of adhesion [57], their measurements exploited the non-parabolic shapes of worn tips, where pull-off force does depend on range of adhesion. In separate experiments also involving AFM pull-off measurements, Jiang et al. have measured a range of adhesion between UNCD and PMMA of 1.5–2.5 nm [20]. Similarly, in nanoindentation experiments adhesive forces were found to act over distances of 1.5–4.5 nm [58]. As mentioned in the introduction, the presence of long-range forces has also been observed by DelRio et al. [18] in adhesion experiments involving silicon micro-cantilevers. The experiments showed contributions to adhesion from distances up to tens of nanometers.

The origin of these large values for range of adhesion is still in dispute. Previously proposed explanations involve electrostatic interactions due to contact charging [59–61], capillary adhesion [62–64], and Casimir forces [18]. First, electrostatic interactions have been proposed as a possible explanation because of the well-known phenomena of contact charging [60, 65–68]. While the detailed physical mechanism is still in discussion [69], the results are a net charge between the two materials that can result in measurable electrostatic interactions. In the present testing, these long-range electrostatic forces would be expected to be seen as measurable forces observed before and after contact. This can manifest as a tilting of the "out-of-contact" region of the force–displacement curve or as an earlier-than-expected snap-into-contact event as the charged sphere attracts the uncharged substrate in a new location. While such long-range interactions *have* been observed in other, unrelated testing where ruby tips were brought into contact with non-conductive substrates, the present substrates were boron-doped for conductivity, and the tester and substrates were electrically grounded to the vacuum chamber. The measured force curves in the present testing were similar to that shown in Fig. 2d, with no interaction forces observed until contact was initiated. Additionally, the operation of a static-reducing ionization bar had no measurable effect on adhesive forces. Therefore, contact charging is not expected to have played a significant role in the present results.

A second common explanation for longer-than expected values for range of adhesion is capillarity. Water bridges across a contact can increase the area of interaction of a rough contact and are known to significantly increase the adhesive force. The relative humidity determines the presence and size of these capillary bridges, which in turn affect the adhesive force. The present testing was carried out in a dry atmosphere (< 1% RH). This is insufficient to eliminate all water from the contact, but will limit its contribution. He et al. [70] showed that, even for hydrophilic surfaces, capillary necks could not form below a relative humidity of 20–40%. Numerical analyses [71] also suggests that capillary formation should not play a role in adhesion at low humidity. Therefore, capillarity is not expected to be the dominant factor in explaining the effect of topography on adhesion.

A third common explanation for large values for range of adhesion is retarded dispersion, or Casimir, interactions. These forces arise due to the finite speed of electromagnetic interactions and typically act over ranges larger than a few nanometers, even up to tens of nm [18]. The present investigation is consistent with these findings, since the large micronewton adhesive forces cannot be explained without considering longer-range interactions. Furthermore, the smoother surfaces show a stronger contribution from these longer-range interactions, while these interactions play a less important role for the rougher surfaces, with fewer, sharper asperities in contact. However, the interaction potential used here does not explicitly account for any specific attraction mechanism. It is an empirical potential that elucidates the strength and length-scale of the interaction. In this case, those parameters are consistent with Casimir forces, but further investigation would be required to conclusively demonstrate the physical origin.

## The Role of Plasticity in Adhesion of These Contacts

To specifically verify that plastic deformation can occur in these contacts, an additional investigation was performed with AFM imaging performed on the ruby sphere (softer material) in the exact location of contact before and after an array test was performed. The standard adhesion test setup does not permit this precise knowledge of test location; therefore, an alternate custom micromechanical test setup was used, with a cantilever based force sensor, but otherwise similar setup. A ruby sphere was polished, pre-imaged in the AFM (Fig. 5b), and then used in an array test of adhesion against an NCD substrate. This testing repeated the 400 measurements in an array of locations from a typical adhesion experiment, but used a preload of 20 μN (the minimum load of the alternate test apparatus). The apex of the sphere was imaged again after the test (Fig. 5c) and the location was matched to the pre-test image. The AFM imaging presented clear evidence of indentations at the tip of the sphere. The indentations were approximately 150 – 300 nm in lateral size, and approximately 2 – 10 nm in depth. These indentations were scattered across the tip of the sphere, with single indents concentrated around a region of multiple overlapping indents.

The numerical modelling (recalculated at a preload of 20 μN) predicted deformations with a depth of approximately 2 nm and edge lengths of approximately 50 nm (Fig. 5a). The measurements are in reasonable agreement with predictions. The region of overlapping indentations makes it difficult to determine the size of a single indentation. However, there do appear to be single indentations scattered around that region. The measured deformation for what appear to be single indents had depths ranging from single nanometers to nearly 10 nm and lateral sizes for measured deformation of approximately 100 – 500 nm. The computed deformations are for a single adhesion test, while the experimental deformations correspond to the cumulative effect of 400 adhesion tests against different contact points. Therefore, the overall scales of deformations compare favorably, and likely indicate that the tallest asperities on the substrates are serving to permanently indent the polished spheres.

### Determining the Most Relevant Length-scales of Roughness

The numerical analysis in this investigation has demonstrated that adhesive interactions act over a length scale of greater than 5 nm, and that plasticity serves to increase contact area above the predictions of rigid or elastic calculations. These two factors may limit the impact of certain length-scales of topography on the total macroscopic adhesion.

To check the influence of different roughness scales, we repeated the pull-off force calculations on a variety of virtual surfaces, with different scales of roughness filtered out or included. This was accomplished by taking advantage of the all-scale measurements performed on the same substrates in Ref. [27] and combining them with the AFM measurements performed here. Since we do not have multi-scale measurements taken in the exact same location, we used the statistics of the random roughness to add smaller- and larger-scale roughness to the measurements. Specifically, we started with an AFM image of the surface of the type shown in Fig. 1, then we superimposed artificially generated roughness that was created using a Fourier-filtering algorithm [72, 73] based on the measured PSD for that particular substrate. Therefore, these virtual surfaces are representative of the true multi-scale topography of each substrate. Then, from these multi-scale "master" surfaces, we filtered out different scales of roughness. Finally, we performed the numerical calculations on each of the filtered surfaces to compute the pull-off force and determine the sensitivity to different scales of roughness. The detailed approach of creating and filtering these surfaces is described in the next paragraph.

To add small-scale roughness, we first stitched the $512 \times 512$ pixel AFM scan using mirror images, leading to a $1024 \times 1024$ periodic topography. This stitched surface was first Fourier interpolated on an $8192 \times 8192$ grid (0.625 nm pixel size) and parts of the spectrum with wavelength smaller than $\lambda_T = 20$ nm were cut out. A randomly rough surface that follows the substrate PSD for wavelengths $\lambda < \lambda_T$ and has a constant roll-off above $\lambda_T$ was added to this interpolated topography. Features below the varying cutoff wavelengths $\lambda_S$ are then filtered out to check their effect on the calculated pull-off force. To add large-scale roughness, the AFM scan was again stitched to create a periodic topography, and this was stitched multiple times to yield an $8192 \times 8192$ (20 μm linear size) grid. Fourier components at wavelengths bigger than $\lambda_T = 1$ μm were cut out. A randomly rough surface with spectrum following the substrate PSD for wavelengths $\lambda > \lambda_T$ and zero below $\lambda_T$ was added to this topography. Features with wavelength above the varying cutoff wavelengths $\lambda_L$ are then filtered out to check how they affected the calculated pull-off force.

Figure 6 shows the change in pull-off that occurs when different length scales of roughness are filtered out. In Fig. 6a, which shows the effect of small-scale roughness, the leftmost datapoints represent the pull-off force computed on the unfiltered surface. As the short-wavelength cutoff (x axis) gets larger, more and more small-scale roughness is removed from the surface. Thus, the surface is perfectly smooth below this cutoff; the rightmost data points approach the pull-off force from a perfectly flat plane. The value of pull-off force remains constant (within 10%) until the cutoff wavelength reaches 43 nm. Removing roughness *above* this size scale has a strong effect on the adhesive force, but removing roughness *below* this size has almost no effect. Figure 6b shows a similar calculation, but now with long-wavelength roughness filtered out; thus, the rightmost datapoint represents a nearly unfiltered surface. As the long-wavelength cutoff decreases, shorter and shorter wavelengths of roughness are removed from the surface, along with all wavelengths above the cutoff. The results show that, with removal of roughness with wavelength above approximately 1.8 microns, the adhesion remains constant (within 10%). Taken together, the two plots in Fig. 6 demonstrate that there is almost no effect on adhesion from roughness with lateral length scales smaller than 43 nm or larger than 1.8 microns. The critical finding of this analysis is that there is a certain band of length scales of roughness, 43 nm to 1.8 microns, that most-strongly affects adhesion in these materials; roughness outside of this band plays a secondary role in adhesion.

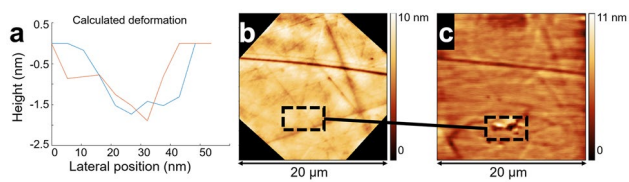

**Fig. 5** The computed deformation for a single adhesion test is shown in panel (**a**) with the red and blue lines representing x and y direction line scans respectively. The actual deformation of the ruby sphere after an array of adhesion tests is measured using AFM images taken at the sphere apex before (**b**) and after (**c**) testing. The images have been precisely located at the apex of the tip where contact took place, and fiducial markers have been used to orient the image. The after-test image confirm the presence of permanent deformation, as is assumed in the numerical modeling, seems to be in order-of-magnitude agreement with what would be expected after 400 tests in different locations, each with the deformation shown in (**a**)

The explanation for this critical band of scales of topography may be different for the large and small scales. The unimportance of large-scale topography is likely linked to the area that is interacting with the sphere. Given a range of adhesion $\rho$, a sphere of diameter $D$ will interact with a flat surface within a disk of radius $r = \sqrt{D\rho}$. For $D = 500$ nm and $\rho = 5.6$ nm we obtain a radius of $r = 1.7$ μm, almost exactly the wavelength above which large-scale topography no longer matters. This shows that macroscopic pull-off forces are strongly affected by finite-size effects, and that the magnitude of pull-off forces will depend strongly on the sphere radius. This also means that the scales of roughness that matter are determined by the macroscopic contact geometry, as long as sphere radius $R$ is much larger than typical scales of the roughness.

For the unimportance of small-scale topography, there are two effects that enhance each other: the large range of adhesion, and the effect of plasticity. The large range of adhesion (5.6 nm), which was determined from the numerical analysis, indicates that topography variations below this scale have a reduced contribution to adhesion. For example, for rigid surfaces with a sinusoidal gap of amplitude 2 nm, a range of adhesion of 0.5 nm would mean that only the contacting peaks contribute to adhesion and the rest of the surface is irrelevant; while a range of adhesion of 5 nm would mean that all portions of the surface are adhering, with only small differences in relative contributions from different locations. An additional explanation is the effect of plasticity. The small-scale roughness has the highest local slope, and thus the highest surface stress. This means that the small-scale roughness will cause deformation, which smooths out these scales earlier than other scales, and reduces their contribution to macroscale adhesion.

## Implications of the Present Findings

The results demonstrated that, for the macroscale adhesion of extremely stiff materials, the very smallest scales of roughness do not determine adhesion. This is in stark contrast to the adhesion of nanoscale contacts of hard materials [19] and to the adhesion of macroscale contacts of soft solids [28], both of which show a critical influence of smallest-scale roughness. While the present work draws on extensive roughness characterization at all scales using stylus profilometry, AFM, and TEM, in the end the AFM-scale roughness data (which covers the critical band of length scales discussed in the prior section) was sufficient to describe adhesion in these contacts. The introduction of smaller-scale roughness, as measured in the TEM, had little influence on the predicted adhesion. This means that parameters like RMS slope and curvature, that are most strongly influenced by the smallest-scale roughness, are less important for these hard-material contacts.

In these measurements, the larger scales of roughness were also less significant. This implies that measurements based on stylus profilometry, which is resolution-limited by the micron-scale radius of the tip, are not sufficient to predict and describe adhesion of these materials. It also implies that a simple scalar parameter such as RMS height is insufficient to determine macroscale adhesion. We look forward to investigating this point further, with the goal to understand the generalizability of this result beyond the current experimental setup.

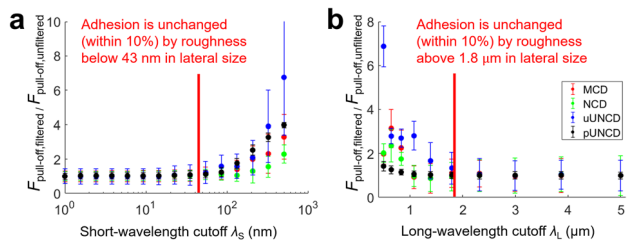

**Fig. 6** The contribution to pull-off force from various length scales can be directly demonstrated by recalculating pull-off force after filtering out small (**a**) and large (**b**) scales of roughness. Specifically, the pull-off force calculated from the filtered surfaces is normalized by the pull-off force calculated from the unfiltered surfaces. In panel (**a**), the x-axis indicates a short-wavelength cutoff, where all roughness below this size scale has been removed. A value near 1 indicates that there is almost no effect on pull-off force of filtering out roughness below that size scale. In panel (**b**), the x-axis indicates a long-wavelength cutoff, where all roughness above this size is removed. Here, a value of 1 indicates no contribution to pull-off force from roughness above that size scale

Another key result of the calculations for hard materials is that adhesion is dominated by the asperities at the very high end of the height distribution. This leads to highly variable extremes in adhesion that can far exceed common predictions based on the average asperity height. The adhesion distributions appear to be log-normal, with a long tail, which strongly impacts the mean adhesion value and leads to rare but significant ultra-high-adhesion events. This has strong implications for real-world applications, such as MEMS devices, which must overcome such surface forces and the threat of stiction. The shape of the measured distributions would suggest that any moving parts should be significantly overdesigned to ensure they can overcome the long-tail events.

Another important finding is that the experimental results were unable to be fit without the inclusion of plasticity. The assumption is that at some length scale, the contact pressure will overcome the hardness of one of the materials. Ruby, in our experiments, is the softer material. Whether this results in plastic flow or fracture, the contact area should evolve to support the preload applied and will be significantly larger than predictions from elastic models. In recent work in both SEM [74] and TEM [51, 52] experiments, plasticity in nanoscale ruby and diamond samples has been reported. Simple experiments were performed to confirm the presence of small-scale plasticity on the ruby tip. These findings are supported by prior work demonstrating connections between nanoscale plasticity and large-scale properties [75]. The nanometer-scale deformation is likely an important factor behind the presence of a small-wavelength cutoff in the roughness that affects adhesion.

The *effective* work of adhesion (that includes the effect of topography) of these surfaces varies by almost two orders of magnitude, from 0.08 to 7.15 mJ/m$^2$. These values for $W_{adh,eff}$ are calculated from the overall sphere geometry using spherical contact-mechanics models. Given the wide variability of the adhesion force between interfaces of identical large-scale geometry, it is not surprising that the effective work of adhesion varies so much. These variations, however, are not explained by simple analytical models, such as those based on a Gaussian distribution of asperity heights, nor those based on a balance between elastic and adhesive energy.

Common (elasto-)adhesion theories balance the elastic energy required for deformation with the interfacial energy (intrinsic work of adhesion) gained during contact [2, 15, 76–78]. In our case, the interfaces are so stiff that the deformation energy vastly exceeds any energy gain from making contact and we expect no pull-off force (or no stickiness [2]) in the "thermodynamic" limit of large surface areas and vanishing range of adhesion. In our case, the pull-off force is then determined by the interfacial stress carried by the intermolecular potential between the two surfaces and we can simply compute it by summing up these stress contributions (as we did in our numerical calculations); or in other words, the interface does not separate like a crack [76]. The explanation for the appreciable pull-off force is tightly linked to the long range (5.6 nm) of interaction extracted from this analysis.

## Conclusions

By combining detailed measurements of topography, thousands of mm-scale adhesion measurements, and numerical integration of an interaction potential, we computed both the intrinsic material parameters governing adhesion as well as the contributions to adhesion from multi-scale topography. The intrinsic work of adhesion between ruby and polycrystalline diamond was found to be 46.3 mJ/m$^2$ while the range of adhesion was 5.6 nm. This large value for range of adhesion, along with the requirement for plasticity in the calculations, leads to a diminished role of small-scale topography on the macroscale adhesion of these hard contacts. While prior work on soft-material adhesion on the same substrates [28] demonstrated the important role of single-digit-nm topography on adhesion, the same is not true for the present measurements of hard-material adhesion. In fact, based on this analysis incorporating plasticity and the large range of adhesion, it has been demonstrated that there is a critical band of length scales of topography—43 nm to 1.8 μm—which plays the most significant role in macroscale adhesion for these hard materials.

**Acknowledgements** We thank Nathaniel Miller for useful discussion and feedback on the manuscript. TDBJ, AG, and LAT acknowledge support from the National Science Foundation under award CMMI-1727378. AS and LP acknowledge funding by the Deutsche Forschungsgemeinschaft under Germany's Excellence Strategy (project EXC-2193/1 – 390951807) and by the European Research Council (Starting Grant 757343). Use of the NanoFabrication and Characterization Facility (NFCF) in the Petersen Institute for Nano Science and Engineering (PINSE) is acknowledged.